\documentclass{article}
\usepackage{float}

\usepackage{arxiv}

\usepackage[utf8]{inputenc} 
\usepackage[T1]{fontenc}    
\usepackage{hyperref}       
\usepackage{url}            
\usepackage{booktabs}       
\usepackage{amsfonts}       
\usepackage{nicefrac}       
\usepackage{microtype}      
\usepackage{lipsum}
\usepackage{graphicx}
\usepackage{amsmath}
\usepackage[]{natbib}
\title{Gaussian Process Regression for Estimating EM Ducting Within the Marine Atmospheric Boundary Layer}

\author{
  Hilarie Sit\\
  Civil and Environmental Engineering\\
  Cornell University\\
  Ithaca, NY 14853 \\
  \texttt{hs764@cornell.edu} \\
   \And
  Christopher J. Earls \\
  Civil and Environmental Engineering\\
  Center for Applied Mathematics\\
  Cornell University\\
  Ithaca, NY 14853 \\
}

\begin{document}
\maketitle

\begin{abstract}
We show that Gaussian process regression (GPR) can be used to infer the electromagnetic (EM) duct height within the marine atmospheric boundary layer (MABL) from sparsely sampled propagation factors within the context of bistatic radars. We use GPR to calculate the posterior predictive distribution on the labels (\textit{i.e.} duct height) from both noise-free and noise-contaminated array of propagation factors. For duct height inference from noise-contaminated propagation factors, we compare a na\"{i}ve approach, utilizing one random sample from the input distribution (\textit{i.e.} disregarding the input noise), with an inverse-variance weighted approach, utilizing a few random samples to estimate the true predictive distribution. The resulting posterior predictive distributions from these two approaches are compared to a ``ground truth'' distribution, which is approximated using a large number of Monte-Carlo samples. The ability of GPR to yield accurate and fast duct height predictions using a few training examples indicates the suitability of the proposed method for real-time applications.
\end{abstract}


\section{Introduction}
Turbulent transport of momentum, moisture, and heat between the ocean and the atmosphere characterizes the marine atmospheric boundary layer (MABL), which is the region of the lower troposphere that directly contacts the ocean surface \citep{Sikora}. Spatiotemporal inhomogeneity in temperature, pressure, and humidity within the MABL can result in anomalous propagation of electromagnetic (EM) waves by altering the refractive index, which is the speed of light in medium relative to that in vacuum, within this region. Rapid decrease in the refractive index with increasing altitude can create a ``trapping'' layer where EM waves are refracted back toward the ocean surface. This phenomenon, called \textit{atmospheric ducting}, can impact performance of radar and communication systems by causing unexpected holes in coverage, inaccurate measurement of elevation angle, and extension of radar horizon \citep{Skolnik}. Thus, it is of great interest to devise real-time methods that can accurately identify and characterize these EM ducts, so that, for instance, radar operators can be informed about their system's expected performance in real-time.

A method for estimating duct characteristics is by calculating the refractive index using direct measurements of atmospheric conditions within the MABL. The refractivity profile, which is related to the refractive index, can be written as an empirical equation parameterized by atmospheric temperature, pressure, and humidity that can be measured via radiosondes or rocketsondes \citep{Bean}. However, direct measurements of atmospheric conditions are necessarily sparse and implementation can be expensive, rendering this method impractical for real-time estimation of MABL refractivity. Other methods for estimating the refractive index include using GPS \citep{Lowry} and LIDAR \citep{Willitsford} measurements. However, these methods are not practical in the context of the duct characterization problem due to its dependence on favorable satellite alignment over the horizon and favorable weather conditions. 

More recently, refractivity from clutter (RFC) methods have gained traction in the literature. RFC methods use radars to measure clutter (\textit{i.e.} backscattered power from the rough ocean surface); thus employing backscattered returns to estimate the refractivity profile. In RFC, the forward solver is called multiple times to predict clutter under certain ducting conditions, and these solutions are compared with new clutter observations to infer refractivity profiles. Such methods frequently use the forward solver and so a full, 3D solution of Maxwell's equations, can be computationally expensive. The Fourier split-step parabolic equation (SSPE) is commonly used as a fast and efficient forward model, however, even SSPE solutions may prove unwieldy for RFC inversions. As a result, efforts have focused on improving efficiency in optimization within the context of the RFC inversion: approaches in the literature include nonlinear least squares \citep{Rogers2000}, matched-field processing approaches \citep{Gerstoft}, Markov-chain Monte Carlo \citep{Yardim}, Markov state-space models \citep{Vasudevan}, particle swarm optimization \citep{Wang}, opposition-based learning \citep{Yang}, and dynamic cuckoo search \citep{Zhang}. A review of RFC methods is available in \cite{Karimian}.

Apart from RFC methods, approaches for constructing simplified forward models using observations obtained by sparsely sampling EM power within the MABL have emerged. \cite{Fountoulakis} utilize blurring operators to approximate effects of the MABL, so that ``just in time'' estimates, for inferring duct parameters, may be obtained by manifold interpolation within a library of sparse proper orthogonal decomposition (POD) modes calculated offline from field observations. These field observations are sparsely sampled along a sinusoidal UAV flight path or a linear rocketsonde flight path. \cite{Gilles} bypass the full forward model by decomposing the governing partial differential equation into few propagating trapped normal modes and solving an easier optimization problem posed on normal mode subspaces. In the context of this method, coverage is sparsely sampled, so as to be consistent with a bistatic case, in which a receiver is located downrange of a hypothetical transmitter.

In prior work, the current authors show that artificial neural networks (ANNs) can accurately and efficiently predict duct height from sparsely measured EM propagation factors \citep{Sit}. Similar to the above methods, the authors simulate coverage diagrams for duct heights of interest and utilize a series of sparse sampling schemes, that are consistent with practical deployment within bistatic contexts, to construct the dataset needed for training and testing. Parameters within the neural networks are then learned by minimizing the empirical risk on the training examples, and model generalizability is evaluated by assessing performance on test examples. While success was achieved in the foregoing work, in general, artificial neural networks are limited by small datasets in the duct characterization problem, as such models may be prone to overfit on the necessarily small training sets.

In the present paper, we propose using Gaussian process regression (GPR) to perform a similar task of predicting duct height from sparsely sampled propagation factors. GPR is a nonparametric, fully-Bayesian method that offers several advantages over other machine learning techniques, such as model interpretability and uncertainty quantification for predictions \citep{Rasmussen}. Using GPR, we tune the prior distribution over functions and subsequently calculate a predictive distribution to estimate duct height from propagation factors obtained via sparse bistatic sampling schemes.

This paper is outlined as follows. We begin by describing the forward model and EM sampling schemes that correspond to different bistatic radar configurations. We then explain Gaussian process regression and its application to this specific duct height prediction task as well as approaches for tackling severely noise-contaminated observations. Finally, we show and discuss the results and limitations of our study.

\section{Forward Model}
EM wave propagation between transmitter and receiver antennas can be modeled as an initial-value problem, using the Fourier split-step parabolic equation (SSPE) algorithm as an approximation to the time-independent Helmholtz wave equation \citep{Ozgun}:

\begin{equation} 
\frac{\partial^2 \varphi}{\partial x^2} + \frac{\partial^2 \varphi}{\partial z^2}  + k^2n(z)^2\varphi = 0
\end{equation}

where $\varphi$ is the horizontal polarization of the electric field, $k$ is the free-space wavenumber, and $n$ is the refractive index. The refractive index, assumed to be height dependent and constant in range, is the ratio of EM propagation velocity in free space to that in the medium. We often define another quantity called \textit{refractivity}, $N = (n - 1)10^6$, to better study the small changes in the refractive index that characterize the MABL and \textit{modified refractivity} to additionally account for the curvature of Earth's surface \citep{Skolnik}. In the presence of evaporation ducts under neutral stability conditions, the modified refractivity profile within the MABL can be modeled as a log-linear curve using the Paulus-Jeske model:

\begin{equation}
M(z) = M_0 + c \left(z - z_d ln\left(\frac{z+z_0}{z_0} \right) \right) 
\end{equation}

where $M_0$ is surface refractivity, $c = 0.13M$-$units/m$ is critical potential refractivity gradient for neutral evaporation ducts, $z_0 = 0.00015m$ is aerodynamic surface roughness of the ocean, and $z_d$ is the duct height \citep{Karimian}. To be consistent with the South China Sea Monsoon Experiments in 1998 \citep{SCSMEX}, we chose $M_0 = 428.89 M\textrm{-units}$ and solve the EM wave equation using SSPE for all considered duct heights, $z_d$, at every half meter between 2-40m in altitude, which encompasses practical evaporation duct height instances \citep{Fountoulakis}.

\subsection{SSPE Solution}
The present work considers a 2D rectangular problem domain with a maximum range (x) of 50km and altitude (z) of 113m, with grid spacings of 40m and 0.1m, respectively. The transmitter antenna is located at $x=0 km$ and generates a horizontally-polarized Gaussian antenna pattern with a radar signal frequency of 9.3 GHz. The SSPE solution is obtained by specifying the initial field at the left boundary and then propagating the field down range by a series of fast Fourier transformations, while satisfying the lower and upper boundary conditions.

We use an adaptation of PETOOL, a SSPE solver developed by \cite{Ozgun}, that specifies a Leontovich surface impedance condition at the lower boundary \citep{Gilles}. The continuity of the tangential components of the electric and magnetic fields in this boundary condition is satisfied by assuming that the ocean free surface at $z = 0m$ is a flat, finite conductor with a homogeneous dielectric constant. This constant is consistent with the thermodynamic conditions of the South China Sea in \cite{SCSMEX} (\textit{i.e.} 100\% humidity at ocean surface, surface temperature of 29.7 $^{\circ}C$, and ocean salinity of 35 ppt.) and can be calculated using the semi-empirical Debye expression \citep{Ryan}: 

\begin{equation}
\epsilon(\omega) = \epsilon_{ir} + \frac{\epsilon_{0} - \epsilon_{ir}}{1-i\omega\tau} + \frac{i\sigma}{\omega\epsilon_{0}}
\end{equation}

\noindent where $\epsilon_{ir} = 4.9$ is the far-infrared dielectric constant of water, $\tau$ is the relaxation time, $\sigma$ is the ionic conductivity, and $\epsilon_0$ is the static dielectric constant of sea water. At infinity, the electric field is decayed to zero in the Sommerfeld radiation condition. Due to the electric field truncation at $z = 113m$, PETOOL approximates this condition at the upper boundary by extending the domain altitude and applying a Hanning window to remove non-physical reflections that result from this truncation. Please refer to Figure 1 for the setup of SSPE.

\begin{figure}[H]
\centering
\includegraphics[width=0.7\linewidth]{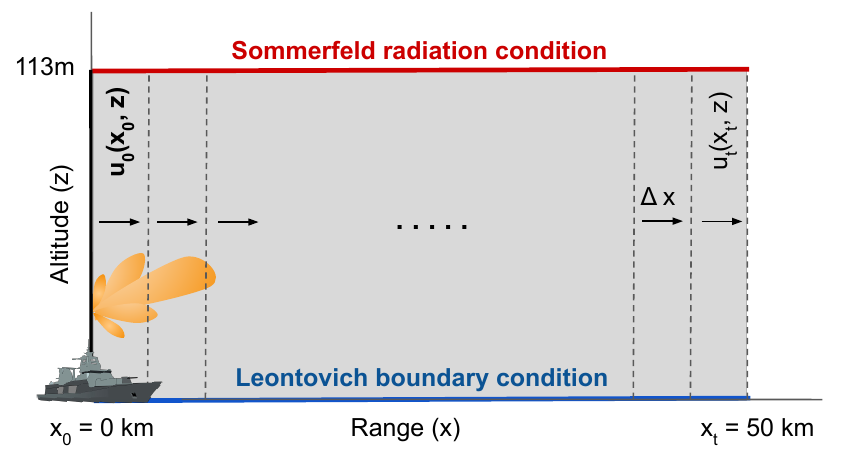}
\caption{\textbf{Problem domain.} Domain of interest and associated boundary and initial conditions.}
\end{figure}

Equipped with the EM wave solution, \textit{propagation factors} within the domain, collectively called the \textit{coverage diagram}, can be calculated \citep{Ozgun}:

\begin{equation}
PF = 20log|u| + 10logx + 10log\lambda
\end{equation}

where $u = exp(-ikx)\varphi(x,z)$ is the reduced amplitude function for the parabolic equation and $\lambda$ is the free-space wavelength. Propagation factors (\textit{\textit{i.e.}} the features of interest in our problem) are defined as the electric field magnitudes scaled with that observed in free space and they account for environmental effects and surface roughness in radar calculations \citep{Ryan}.

\subsection{EM Sampling Schemes}
Propagation factors are sparsely sampled from within the coverage diagram to provide a practical context for collecting bistatic radar measurements. We use three EM sampling schemes to construct three separate datasets, each consisting of an array of propagation factors (observation/input) for every considered duct height (label):

\textbf{Case 1: Stationary transmitter and moving receiver.} We consider a stationary transmitter with antenna height of $h = 10m$ and angle of $\alpha = 0^{\circ}$, with respect to the horizontal, and a moving receiver attached to a \textit{rocketsonde} (\textit{i.e.} solid rocket propelled sensing system). This rocketsonde-receiver sampling system (RRSS) is flown at a constant altitude of $y=21m$, and samples 250 evenly spaced propagation factors along a horizontal line between 5-15km in range. Please refer to Figure 2 for a depiction of this sampling path. This flight trajectory was arrived at as a result of an exhaustive sensitivity study aimed at uncovering short, flat sampling paths that lead to successful evaporation duct height inversions. 

\begin{figure}[H]
\centering
\includegraphics[width=0.8\linewidth]{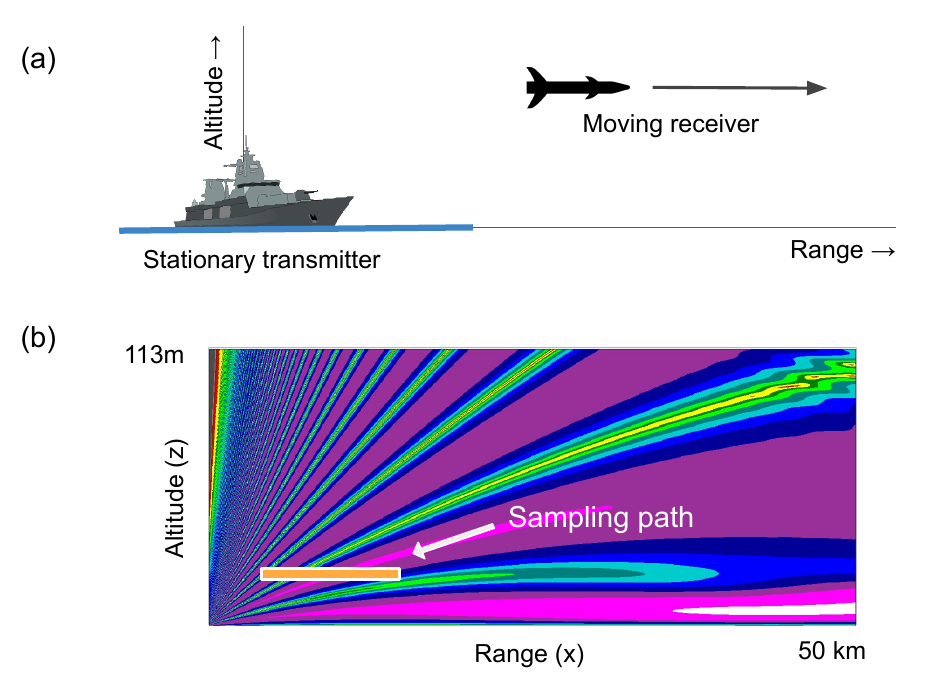}
\caption{\textbf{EM Sampling Scheme 1.} Path is consistent with \textbf{a.} a stationary transmitter and moving receiver (RRSS), and is \textbf{b.} shown on representative coverage diagram. }
\end{figure}

\textbf{Case 2: Deterministically rocking transmitter and stationary receivers.} We consider a rocking transmitter and stationary receivers mounted onto a vertical tower located at $x=50km$ downrange. These receivers sample 30 equally spaced propagation factors between 0-30m in altitude. Please refer to Figure 3 for a depiction of this sampling scheme. The rocking is meant to be more closely aligned with the notion of a floating transmitter. To mimic arbitrary rolling of $\pm0.5^{\circ}$ and heaving of $10m$ of the transmitter antenna, six observations are collected using discrete combinations of transmitter heights $h = 20m, 30m$ and antenna angles $\alpha = -0.5^{\circ}, 0^{\circ}, 0.5^{\circ}$, concatenated within an array of propagation factor measurements.

\begin{figure}[H]
\centering
\includegraphics[width=0.8\linewidth]{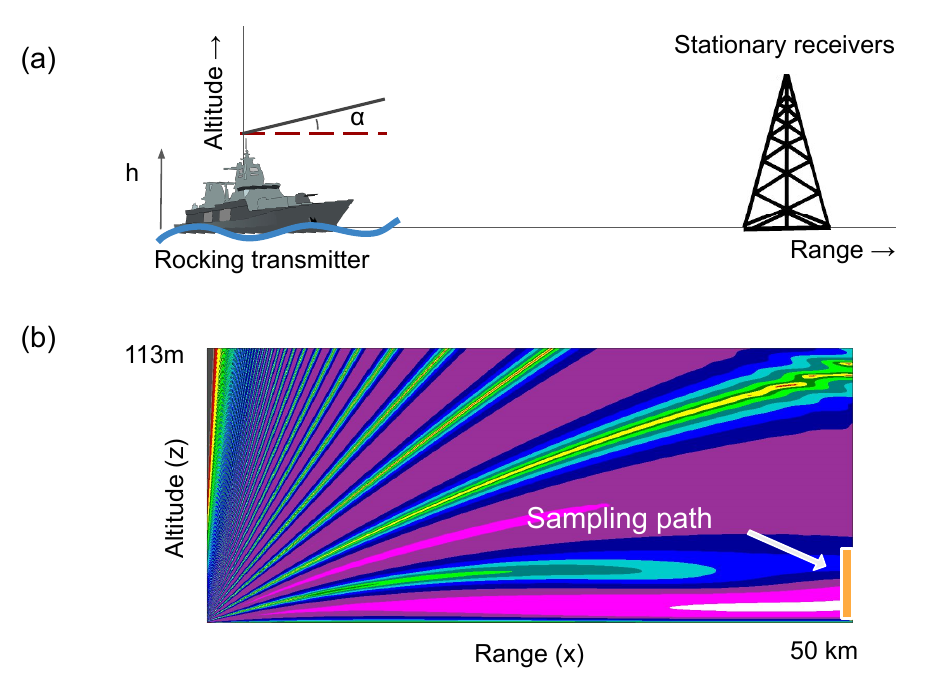}
\caption{\textbf{EM Sampling Scheme 2.} Path is consistent with \textbf{a.} a rolling and heaving transmitter and stationary receivers mounted on a tower, and is \textbf{b.} shown on representative coverage diagram.}
\end{figure}

\textbf{Case 3: Stochastically rocking transmitter and stationary receivers.} Using the same sampling scheme as Case 2, to mimic stochastic rolling and heaving of the transmitter antenna, five observations are collected by sampling uniformly for transmitter antenna heights $h \sim U[20m, 30m]$ and antenna angles $\alpha \sim U[-0.5^{\circ}, 0.5^{\circ}]$, so as to form another array of measured propagation factors. This configuration leads to less of a defined pattern in the dataset compared to the deterministic configuration.

\subsection{Severe Sensor Noise}
To create surrogate experimental results, we contaminate our propagation factor measurements with ``severe'' electronic sensor noise using either additive Gaussian white noise or pink noise. Noise is generated using MATLAB's \texttt{dsp.colorednoise} function, which generates colored noise with power spectral density of $1/|f|^\beta$, where $\beta$ corresponds to the inverse frequency power (\textit{e.g.} $\beta = 0$ for Gaussian white noise, $\beta = 1$ for pink noise), and then applies a color filter if $\beta \neq 0$. The noise is scaled using $\sigma = 0.1||x||_\infty$, where sigma is 0.1 times the absolute value of the largest propagation factor from a given measurement array for white noise and $\sigma /1.6$ for pink noise. The factor 1.6 has been arrived at through a statistical analysis of pink and white noise contaminated signals, so as to result in second moment properties (variance) that are the same between the two, different noise models.
 
\section{Gaussian Process Regression}
Gaussian process regression (GPR) is a nonparametric fully-Bayesian method that performs inference in the space of assumed (prior) functions to ultimately arrive at the posterior predictive distributions at points where we would like predictions. The joint prior distribution is specified as a Gaussian process, which is a collection of an uncountably infinite set of random variables such that any finite subset has a multivariate Gaussian distribution that can be specified using only the mean function, $m(x)$, and covariance function, $k(x, x')$ \citep{Rasmussen}:

\begin{equation} f(x) \sim \mathcal{GP}(m(x), k(x, x')) \end{equation}

If observation labels (on training points) are contaminated with independent and identically distributed (\textit{i.i.d}) additive Gaussian white noise $\epsilon_y \sim \mathcal{N}(0, \sigma_n^2)$, its prior joint distribution can still be a Gaussian process, because the sum of two independent random variables from Gaussian distributions forms another Gaussian PDF that has mean and variance that are the sums of the two distributions. Thus, for noisy labels $y$, we can assume a Gaussian process with the following mean and covariance functions \citep{Rasmussen}:

\begin{equation} y = f(x) + \epsilon_y \end{equation}
\begin{equation} y \sim \mathcal{GP} (m(x), k(x,x')+\delta_{ij}\sigma_n^2)  \end{equation}

where the noise variance, $\sigma_n^2$, is set at $0.1$, to correspond to the grid spacing in the altitude direction of our PETOOL model (as a way to encode ambiguity in the vertical resolution). The functions evaluated on training and test sets are thus, jointly Gaussian:

\begin{equation} 
\begin{bmatrix}
	y \\
	f_\ast
\end{bmatrix} \sim \mathcal{N} \bigg( \begin{bmatrix}
	\mu \\
	\mu_\ast
\end{bmatrix} , 
\begin{bmatrix}
	K(X,X) + \sigma^2_n I & K(X,X_\ast) \\
	K(X_\ast, X) & K(X_\ast,X_\ast)
\end{bmatrix} \bigg)
\end{equation}

where $y$ is the vector containing the noisy training labels, $f_\ast$ is the vector of test predictions, $\mu$ is the vector of training label means, $\mu_\ast$ is the vector of test label means, $X$ are the training observations, $X_\ast$ are the test observations, and $K$ is the covariance kernel that can be obtained by evaluating the covariance function, k, at input pairs. Covariance functions are chosen by the analyst and their associated ``hyperparameters" are tuned by maximizing the log marginal likelihood on the training observations. The posterior predictive distributions for new observations can then be calculated by conditioning the joint prior distribution on the data.

\subsection{Covariance Kernel function}
The covariance kernel function specifies the point-wise similarity within the GP and acts as a constraint on the distribution of functions in Gaussian process regression. A popular covariance kernel function is the squared-exponential (SE) kernel function \citep{Rasmussen}:
\begin{equation} k_{SE}(x, x';\theta) = exp\bigg(-\frac{1}{2\ell^2}(x-x')^2 \bigg) \end{equation}
SE is stationary and infinitely differentiable \citep{Duvenaud}. Within the family of kernels are tunable ``hyperparameters", $\theta$, that further define the distribution of possible functions. The characteristic lengthscale, $\ell$, in SE adjusts the smoothness of the functions (\textit{i.e.} the longer the $\ell$, the more smooth the functions, since a large $\ell$ indicates long range, point-wise similarity in the GP prior) \citep{Rasmussen}. Other common covariance kernel functions include constant, periodic, linear, and Mat\'ern kernel functions. More expressive functions can also be constructed by summing and multiplying these kernel functions. In the present work, we use the product of the constant kernel function and the SE kernel function, as our covariance kernel  function, so that the constant kernel function scales the magnitude of the SE kernel function, to encode the signal variance:

\begin{equation}
k(x, x';\theta) = k_{C}(x, x';\theta)*k_{SE}(x, x';\theta) = \sigma_f^2 \ exp\bigg(-\frac{1}{2\ell^2}(x-x')^2 \bigg)
\end{equation}

where $x$ are the training observations, $x'$ are the test observations, and $\sigma_f^2$ is the tunable signal variance. The initial value of the signal variance is specified as $1.0$ with bounds of (1e-1, 1e3) for case 1 and bounds of (1e-1, 5e1) for cases 2 and 3. The initial lengthscale of the SE kernel is specified as 10.0 with bounds of (1e-3, 1e3). The values indicated were initial values that serve as a point of departure for a numerical optimization algorithm to arrive at appropriate values (vis-\'a-vis training data) via maximization of the log likelihood function.

\subsection{Model Selection}
Model selection for GPR, typically referred to as model ``training", is the process of selecting the family of covariance functions, and setting the free hyperparameters (\textit{e.g.} characteristic lengthscale $\ell$ in squared exponential functions). A common approach is to use an optimization algorithm to find hyperparameters that maximizes the log marginal likelihood. Marginal likelihood is the probability of the data, given model parameters (\textit{i.e.} marginalize out the model parameters); thus, it inherently encodes a tradeoff between model complexity and fit \citep{Rasmussen}: 

\begin{equation}
p(y|X) = \int P(y|f,X)p(f|X)df
\end{equation}
\begin{equation}
log p(y|X) = -\frac{1}{2}y^T(K+\sigma_n^2 I)^{-1}y - \frac{1}{2} log|K+\sigma_n^2 I| - \frac{n}{2}log2\pi
\end{equation}

where $y$ are the labels, $X$ are the observations, and $K$ is the covariance matrix emanating from the kernel function. When maximizing Eqn. 12, gradient-based optimization is preferred, because the inversion of the covariance matrix is performed only once and partial gradients of the log marginal likelihood with respect to the hyperparameters can easily be calculated \citep{Rasmussen}. Typically, inversion of this matrix within the calculation of the marginal likelihood is performed using Cholesky factorization for numerical stability. We use a computer memory efficient, quasi-Newton iterative method for bound-constrained optimization called L-BFGS-B within scikit-learn \citep{Pedregosa}. The log likelihood is typically non-convex and as result, optimization is allowed to be restarted $n=10$ times with different initializations, so as to identify reasonable parameters for the Gaussian process.

\subsection{Prediction and Evaluation}
The joint probability distribution on the training points and the test points is a multivariate Gaussian and thus, when conditioned on the observations, the resulting distribution, called the posterior predictive distribution, is also a multivariate Gaussian \citep{Rasmussen}:

\begin{equation} f_\ast|X,y,X_\ast \sim \mathcal{N}(\bar{f_\ast}, cov(f_\ast)) \end{equation} 
\begin{equation}  \bar{f_\ast} = \mu_\ast + K(X_\ast, X)[K(X,X)+\sigma^2_n I]^{-1}(y-\mu) \end{equation} 
\begin{equation} cov(f_\ast) = K(X_\ast, X_\ast)-K(X_\ast, X)[K(X,X)+\sigma^2_n I]^{-1}K(X, X_\ast) \end{equation} 

As a result, GPR is able to provide a prediction, $\bar{f}_{\ast}$, which is the expected value from all possible functions, and its variance, which can be taken from the diagonal of the covariance matrix, $cov(f_\ast)$, for every desired test point. Calculation of the prediction and variance requires the inversion of the resulting covariance matrix, and again, for numerical stability, Cholesky factorization is often used.

Within the present work, GPR model predictions (\textit{i.e.} duct heights) are evaluated using the mean squared error (MSE) metric, which measures the average squared difference between the predictions and labels.

\begin{equation} 
MSE =\frac{1}{n}\sum_{i=1}^n (\bar{f}_{\ast_i}-y_i)^2 
\end{equation} 

where $n$ is the number of test points (where predictions are made), $\bar{f}_{\ast}$ are the model predictions (mean predicted duct heights), and $y$ are the true labels (\textit{i.e.} actual duct heights). 

\subsection{Observation Noise}
Unlike noise on the labels, which can be easily incorporated into the Gaussian process, noise in the observations (\textit{i.e.} inputs) is problematic. To obtain a prediction for a noise-contaminated input, we need to marginalize out the input distribution from the posterior predictive distribution, which results in a non-Gaussian distribution \citep{Girard}:

\begin{equation} x = u + \epsilon_x \end{equation}
\begin{equation}
p(y|X, f, u, \Sigma_x)  = \int p(y|X, f, x) p(x| u, \Sigma_x)dx
\end{equation}

where $u$ are the ground truth observations and $\Sigma_x$ is a diagonal matrix containing the variance of the \textit{i.i.d.} Gaussian white noise, $\epsilon_x$, on the observation. Several approaches can be used to numerically or analytically approximate the integral (Eqn. 18). A numerical approximation can be obtained using Monte-Carlo sampling, which involves sampling the input distribution and mixing their Gaussian posterior predictive distributions to obtain a numerical estimation of the actual posterior predictive distribution \citep{Girard}: 

\begin{equation}
p(y|X, f, u, \Sigma_x) \simeq \sum^T_{t=1} w_t p(y|X, f, x_t) = \sum^T_{t=1} w_t \ \mathcal{N} (\mu(x_t), cov(x_t))
\end{equation}

where T is the number of samples, $w_t$ is the mixing proportion assigned the Gaussian (note: $\sum^T_{t=1} w_t = 1$), and $x_t$ is a sampled input. For a given $x_t$, we compute the mean, $\mu(x_t)$, and covariance, $cov(x_t)$, of the Gaussian posterior predictive distribution. The resulting distribution from the mixture of Gaussians might not be Gaussian, but the mean and variance of the mixture distribution can still be calculated. The mean is the weighted mean of the means of the Gaussians, and the variance is the weighted variances of the Gaussians with additional terms that account for the weighted dispersion of the means of the Gaussians \citep{Trailovic}:

\begin{equation} 
\mu_{mixture} = \sum^T_{t=1} w_t \mu(x_t)
\end{equation}
\begin{equation} 
\sigma^2_{mixture} = \sum^T_{t=1} w_t (\sigma^2(x_t) + \mu(x_t)^2) - \mu_{mixture}^2
\end{equation}

The prediction can be approximated using the mean, and its 95\% confidence interval can be generated using the variance (Eqn. 21). The desired true distribution can be approached with this approximate distribution by increasing the number of samples; thus we choose large T=1000 with equal mixing proportions $w_t = \frac{1}{T}$ to estimate a ``ground truth'' distribution. 

\begin{figure}[H]
\centering
\includegraphics[scale=0.8]{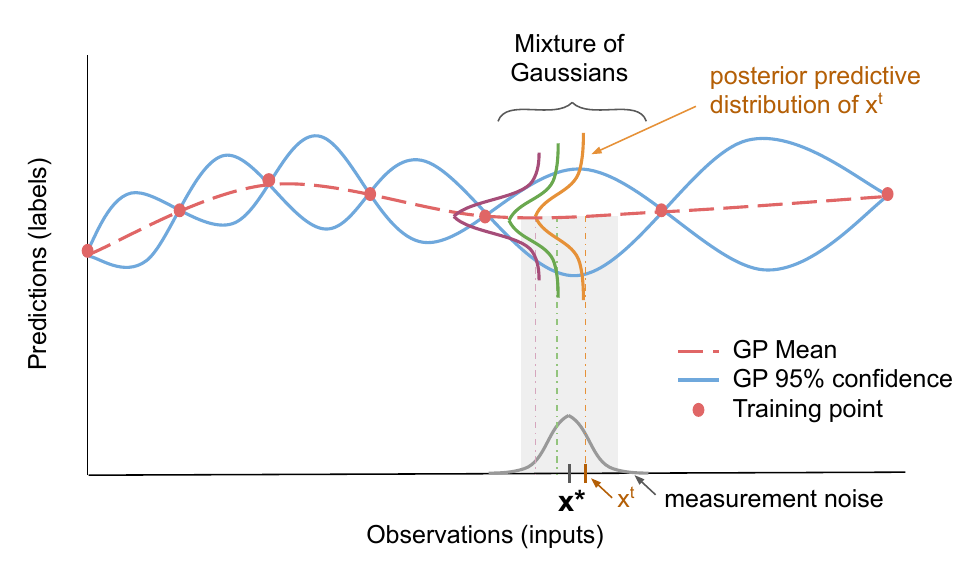}
\caption{\textbf{Monte Carlo sampling of noisy observations.} A numerical estimation method that samples from the test point distribution and mixes the Gaussian posterior predictive distributions (on the labels) obtained for each sample, $x^t$, leading to an approximate posterior predictive distribution for $x^*$. ($x^*$ is shown for a one-dimensional case for simplicity.)}
\end{figure}

While other methods exist for arriving at the mixing proportions, we propose an inverse-variance weighting scheme (Eqn. 22) for use when the number of observations is small. In this way, the Gaussian with the smallest variance can be given the highest mixing proportion, $w_t$:

\begin{equation}
w_t = \frac{1/ \sigma^2_t}{\sum^T_{t=1}1/ \sigma^2_t} 
\end{equation} 

Any of the techniques discussed above can then be applied broadly when we want predictions from an observation contaminated with Gaussian noise. For application, we can assume that a precise sensor is used for obtaining training observations, and the extreme electronic sensor noise is only present on the testing observations.

\section{Results and Discussion}
We use Gaussian process regression (GPR) to predict the duct height from both noise-free and noise-contaminated propagation factors. The mean squared error, compared to the true label, is calculated on all predictions as a metric of model performance. Table 1 shows this metric calculated on (a) noise-free predictions as well as (b) white-noise and (c) pink noise-contaminated predictions obtained from different approaches discussed in Section 3.4. For reference, root mean square error (RMSE) of 0.5m, which roughly corresponds to an MSE of 2.50e-1 represents a reasonable practical ambiguity. Some variability can be expected due to random train/test splits and in the optimization of hyperparameters.

\subsection{Effect of Train to Test Ratio}
To observe GPR performance on even smaller training datasets, we vary the ratio of training points to testing points, from 80/20 down to 50/50. The end points of the dataset, 2m and 40m, are included in the training set, so that the model does not need to extrapolate at those points during inference. As can be seen in Table 1a, model performance is proportional to the train to test ratio. This observation is in line with the fact that a higher number of training points allows the GP to better constrain the prior distribution based on the training data before inference. When the ratio decreases, performance of GPR on cases 2 and 3 degrades below the ideal threshold. Thus, it is recommended to collect data that corresponds to the sampling scheme in case 1, which maintains favorable performance even at a ratio of 50/50.

\begin{table}
\centering
\caption{MSE for Gaussian process regression predictions compared to true label}
\textbf{(a) Noise-free observations} \\ 
\vspace{0.2cm}
\begin{tabular}{l | c c c c}
\hline
Train/Test Ratio & Case 1 & Case 2 & Case 3\\ 
\hline
80/20 & 4.92E-04 & 5.24E-02 & 1.08E-01\\
70/30 & 4.66E-04 & 3.78E-01 & 2.62E-01\\
60/40 & 4.05E-04 & 5.75E-01 & 4.38E-01\\
50/50 & 9.06E-03 & 2.03E+00 & 2.19E+00\\
\end{tabular} \\
\vspace{0.5cm}
\textbf{(b) White Noise-contaminated observations} \\
\vspace{0.2cm}
\begin{tabular}{l c | c c c c | c c }
\hline
& Train/Test Ratio & MC GT$^a$ & Na\"{i}ve & IVW5$^b$ & IVW10$^c$ &\% Imp$^b$ & \% Imp$^c$ \\ 
\hline 
\textbf{Case 1} & 80/20 & 7.81E-03 & 1.28E+00 & 2.15E-01 & 8.97E-02 & 83.2 & 93.0\\
& 70/30 & 1.13E-02 & 1.31E+00 & 2.41E-01 & 8.29E-02 & 81.5 & 93.7\\
& 60/40 & 1.81E-02 & 1.24E+00 & 2.68E-01 & 8.53E-02 & 78.4	 & 93.1\\
& 50/50 & 5.63E-02 & 1.12E+00 & 2.69E-01 & 1.09E-01 & 75.9	 & 90.2\\
\hline
\textbf{Case 2} & 80/20 & 1.48E-01 & 5.37E-01 & 1.84E-01 & 1.29E-01 & 65.8 & 76.0\\
& 70/30 & 7.07E-01 & 1.18E+00 & 7.59E-01 & 6.53E-01 & 35.5 & 44.6\\
& 60/40 & 9.16E-01 & 1.47E+00 & 1.01E+00 & 9.06E-01 & 31.4 & 38.4\\
& 50/50 & 2.70E+00 & 2.95E+00 & 2.83E+00 & 2.63E+00	& 4.18 & 11.0\\
\hline
\textbf{Case 3} & 80/20 & 1.52E-01 & 4.52E-01 & 1.72E-01 & 1.89E-01 & 61.9 & 58.3\\
& 70/30 & 4.07E-01 & 8.64E-01 & 5.11E-01 & 4.93E-01 & 40.8 & 43.0\\
& 60/40 & 9.39E-01 & 1.12E+00 & 9.82E-01 & 1.07E+00 & 12.6 & 5.20\\
& 50/50 & 3.31E+00 & 3.32E+00 & 3.43E+00 & 3.43E+00 & -3.18 & -3.11\\
\hline
\end{tabular} \\
\vspace{0.5cm}
\textbf{(c) Pink Noise-contaminated observations} \\
\vspace{0.2cm}
\begin{tabular}{l c | c c c c | c c }
\hline
& Train/Test Ratio & MC GT$^a$ & Na\"{i}ve & IVW5$^b$ & IVW10$^c$ &\% Imp$^b$ & \% Imp$^c$ \\ 
\hline 
\textbf{Case 1} & 80/20 & 1.22E-02 & 3.28E+00 & 6.83E-01 & 2.06E-01 & 79.2 & 93.7\\
& 70/30 & 1.72E-02 & 2.82E+00 & 5.70E-01 & 2.14E-01 & 79.8	 & 92.4\\
& 60/40 & 2.52E-02 & 2.73E+00 & 6.16E-01 & 2.82E-01 & 77.5	 & 89.7\\
& 50/50 & 6.77E-02 & 2.61E+00 & 5.55E-01 & 2.59E-01 & 78.7	 & 90.1\\
\hline
\textbf{Case 2} & 80/20 & 2.05E-01 & 3.27E+00 & 1.91E+00 & 3.91E-01 & 41.4 & 88.0\\
& 70/30 & 8.07E-01 & 5.91E+00 & 2.26E+00 & 1.69E+00	& 61.8 & 71.5\\
& 60/40 & 1.05E+00 & 6.95E+00 & 2.67E+00 & 2.20E+00 & 61.7 & 68.4\\
& 50/50 & 2.87E+00	& 8.16E+00 & 4.64E+00 & 4.47E+00	 & 43.1 & 45.3\\
\hline
\textbf{Case 3} & 80/20 & 2.32E-01 & 5.60E+00 & 1.22E+00 & 1.36E+00 & 78.2 & 75.8\\
& 70/30 & 5.49E-01 & 6.89E+00 & 2.44E+00 & 2.12E+00 & 64.5 & 69.2\\
& 60/40 & 1.27E+00 & 6.28E+00 & 3.61E+00 & 2.35E+00 & 42.5 & 62.6\\
& 50/50 & 3.87E+00 &8.16E+00 & 5.02E+00 & 4.12E+00 & 38.5 & 49.5\\
\hline
\end{tabular}
\\ $^a$MC GT - Monte Carlo Ground Truth
\\ $^b$IVW5 - Inverse-variance weighting approach for 5 samples
\\ $^c$IVW10 - Inverse-variance weighting approach for 10 samples
\end{table}

\subsection{Effect of Observation Noise}
In the present paper, the model is trained solely on noise-free observations and used to estimate duct heights from observations with significant colored noise contamination, which is unknown to the model. We consider both additive Gaussian white noise and non-Gaussian pink noise to observe the generalizability of GPR techniques on different forms of noise contamination. 

For these noise-contaminated propagation factors, we estimate the true posterior predictive distribution by calculating a numerical approximation using 1000 Monte-Carlo samples (approach shown in Fig. 4). From this, we obtain a ``ground-truth'' distribution, whose mean is the prediction from the noise-free measurement and variance is inflated to account for the input noise, for comparison with the \textit{na\"{i}ve approach} and the \textit{inverse-variance weighting approach}. In the na\"{i}ve approach, a single noise-contaminated measurement is assumed to be deterministic, and the posterior predictive distribution is calculated as if the measurement were a noise-free measurement. In the inverse-variance weighted approach, detailed in Section 3.4, five and ten mixing components are used with inverse-variance weighted mixing proportions (Eqn. 22). MSE on the predictions from the na\"{i}ve approach and inverse-variance approach are compared to the ``ground truth'' duct height, shown in Table 2.

\begin{table}[H]
\centering
\caption{MSE for Gaussian process regression predictions (from na\"{i}ve approach and inverse-variance weighting, IVW, approach) compared to mean of ``ground truth'' distribution}
\vspace{0.5cm}
\textbf{(a) White Noise-contaminated observations} \\
\vspace{0.2cm}
\begin{tabular}{l c | c c c | c c }
\hline
& Train/Test Ratio & Na\"{i}ve & IVW5$^a$ & IVW10$^b$ &\% Imp$^a$ & \% Imp$^b$ \\ 
\hline 
\textbf{Case 1} & 80/20 & 1.21E+00 & 1.71E-01 & 6.29E-02 & 85.9 & 94.8\\
& 70/30 & 1.28E+00 & 2.21E-01 & 6.36E-02 & 82.7 & 95.0\\
& 60/40 & 1.18E+00 & 2.21E-01 & 6.00E-02 & 81.3 & 94.9\\
& 50/50 & 1.04E+00 & 1.94E-01 & 6.14E-02 & 81.4 & 94.1\\
\hline
\textbf{Case 2} & 80/20 & 2.94E-01 & 4.66E-02 & 4.15E-02 & 84.1 & 85.9\\
& 70/30 & 3.89E-01 & 3.34E-02 & 3.13E-02 & 91.4 & 92.0\\
& 60/40 & 3.62E-01 & 4.15E-02 & 2.74E-02 & 88.5 & 92.4\\
& 50/50 & 2.47E-01 & 3.81E-02	 & 2.28E-02 & 84.6 & 90.8\\
\hline
\textbf{Case 3} & 80/20 & 4.07E-01 & 7.26E-02 & 2.92E-02 & 82.2 & 92.8\\
& 70/30 & 3.59E-01 & 6.92E-02 & 3.77E-02 & 80.7 & 89.5\\
& 60/40 & 3.88E-01 & 6.98E-02 & 3.24E-02 & 82.0 & 91.6\\
& 50/50 & 3.44E-01 & 6.91E-02	 & 2.80E-02 & 79.9 & 91.8\\
\hline
\end{tabular} 
\vspace{0.5cm}

\textbf{(b) Pink Noise-contaminated observations} \\
\vspace{0.2cm}
\begin{tabular}{l c | c c c | c c }
\hline
& Train/Test Ratio & Na\"{i}ve & IVW5$^a$ & IVW10$^b$ &\% Imp$^a$ & \% Imp$^b$ \\ 
\hline 
\textbf{Case 1} & 80/20 & 3.01E+00 & 5.96E-01 & 1.69E-01 & 80.2 & 94.4\\
& 70/30 & 2.57E+00 & 5.32E-01 & 2.03E-01 & 79.3 & 92.1\\
& 60/40 & 2.54E+00 & 5.61E-01 & 2.63E-01 & 77.9 & 89.6\\
& 50/50 & 2.55E+00 & 4.62E-01 & 2.35E-01 & 81.9 & 90.8\\
\hline
\textbf{Case 2} & 80/20 & 2.88E+00 & 1.88E+00 & 3.41E-01 & 34.7 & 88.1\\
& 70/30 & 4.29E+00 & 1.39E+00 & 4.95E-01 & 67.6 & 88.5\\
& 60/40 & 4.95E+00 & 1.30E+00 & 6.11E-01 & 73.8 & 87.7\\
& 50/50 & 4.82E+00 & 1.42E+00 & 6.33E-01 & 70.5 & 86.9\\
\hline
\textbf{Case 3} & 80/20 & 6.22E+00 & 1.12E+00 & 1.01E+00 & 82.0 & 83.7\\
& 70/30 & 6.46E+00 & 2.09E+00 & 1.38E+00 & 67.7 & 78.6\\
& 60/40 & 6.16E+00	& 2.98E+00 & 1.38E+00 & 51.7 & 77.6\\
& 50/50 & 6.01E+00 & 2.75E+00 & 1.31E+00 & 54.2 & 78.2\\
\hline
\end{tabular} 
\\ $^a$IVW5 - Inverse-variance weighting approach for 5 samples
\\ $^b$IVW10 - Inverse-variance weighting approach for 10 samples
\end{table}

The inverse-variance weighted approach with five samples offers a large improvement in comparison to the na\"{i}ve approach by decreasing the MSE to between $79.9\%$ to $91.4\%$ for white noise and $34.7\%$ to $82.0\%$ for pink noise, as can be seen in Table 2. Predictions on the pink noise-contaminated observations are slightly worse than those on the white noise-contaminated observations and thus benefit from utilizing more samples in this method. Further improvements can be seen when ten samples are used as the MSE decreases to between $85.9\%$ to $95.0\%$ for white noise and $77.6\%$ to $94.4\%$ for pink noise. However, it is important to note that the performance of this method relies on the ability of the model to get accurate predictions on the noise-free observations, and subsequently, for the ``ground truth''. Performance degrades when this is not true, as can be seen in the last two columns of Table 1 where the improvement in MSE, calculated on the labels, of the inverse-variance weighting approach decreases as the train to test ratio decreases.

Unexpected variance inflation in the presence of noise can be attributed to the high dimensionality of the array of propagation factors (\textit{i.e} adding noise to every dimension can knock the test point further away from the training points in space). As can be seen in Table 3, the average Euclidean distance between the test inputs and their closest training input for the noise-contaminated test inputs is between 1.22 and 2.73 times further away. As can be seen in Figure 6, the increase in variance from using the inverse-variance weighted method made labels be within the 95\% confidence interval of the prediction. The inverse-variance weighted method thus can offer an improvement without the computational cost of MC sampling with 1000 samples.

\begin{table}
\centering
\caption{Average Euclidean distance between test point and its closest training point.}
\vspace{0.2cm}
\begin{tabular}{l c | c c c}
\hline
 & Train/Test Ratio & Noise-free & White Noise & Pink Noise \\
\hline
\textbf{Case 1} & 80/20 & 12.2 & 32.5 (2.66$\times$) & 30.6 (2.50$\times$) \\
& 70/30 & 11.9 & 32.5 (2.73$\times$) & 30.4 (2.55$\times$)\\
& 60/40 & 12.4 & 33.9 (2.73$\times$) & 31.1 (2.50$\times$)\\
& 50/50 & 14.2 & 35.1 (2.47$\times$) & 33.0 (2.32$\times$)\\
\hline
\textbf{Case 2} & 80/20 & 26.6 & 34.5 (1.30$\times$) & 34.6 (1.30$\times$)\\
& 70/30 & 23.0 & 37.5 (1.25$\times$) & 37.9 (1.26$\times$)\\
& 60/40 & 29.8 & 37.1 (1.25$\times$) & 38.1 (1.28$\times$)\\
& 50/50 & 32.0 & 39.2 (1.23$\times$) & 40.0 (1.25$\times$)\\
\hline
\textbf{Case 3} & 80/20 & 27.0 & 33.6 (1.25$\times$) & 33.3 (1.24$\times$)\\
& 70/30 & 28.3 & 34.7 (1.23$\times$) & 34.6 (1.22$\times$)\\
& 60/40 & 28.3 & 35.7 (1.26$\times$) & 35.5 (1.25$\times$) \\
& 50/50 & 30.1 & 37.2 (1.23$\times$) & 36.8 (1.22$\times$) \\
\hline
\end{tabular} \\
\end{table}

\begin{figure}
\centering
\includegraphics[scale=1.1]{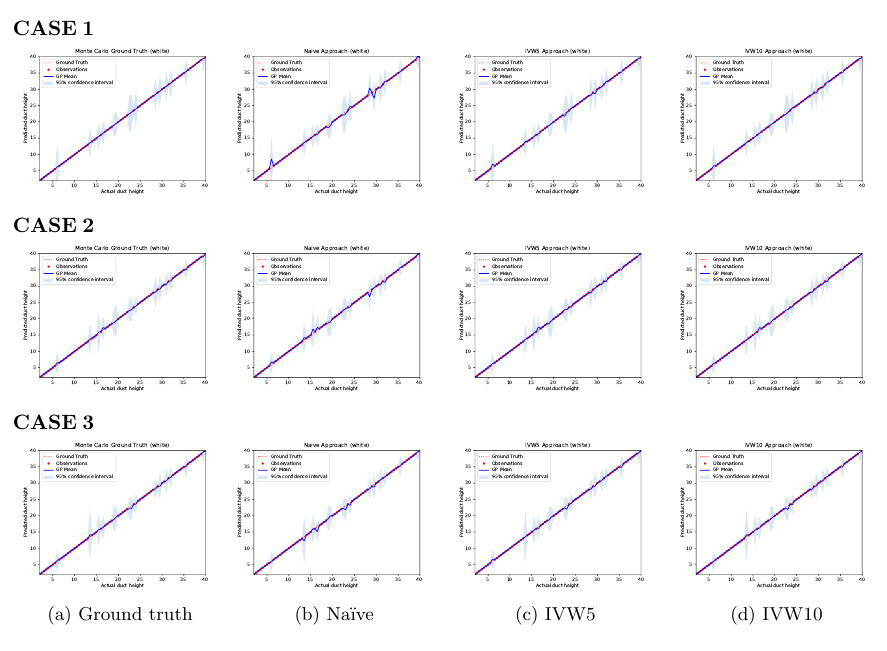}
\caption{GPR results for white noise contamination with 80/20 train to test ratio}
\end{figure}

\begin{figure}[H]
\centering
\includegraphics[scale=1.1]{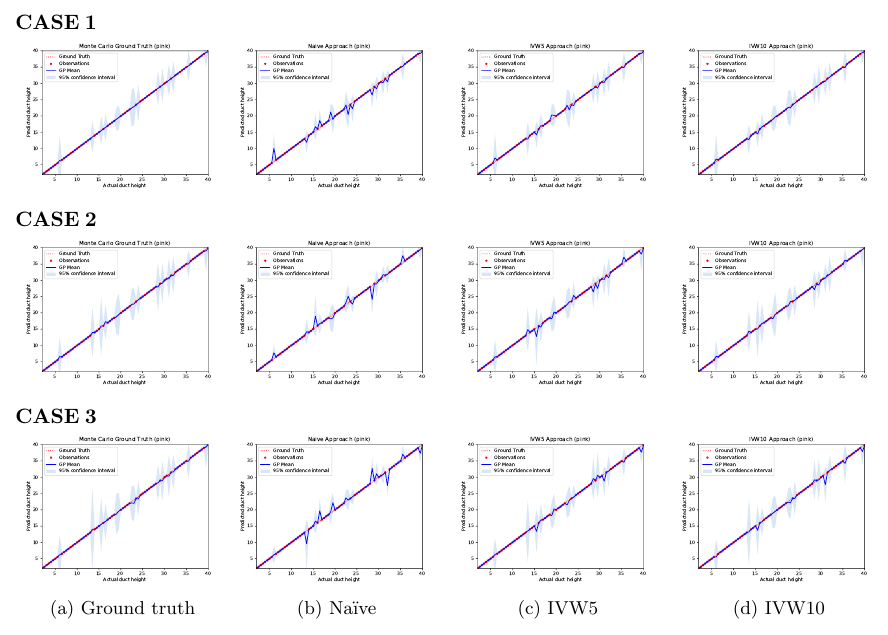}
\caption{GPR results for pink noise contamination with 80/20 train to test ratio}
\end{figure}

\subsection{Timing}
Both training and inference are performed using 4-cores of Intel i5 microprocessor having a clock speed of 2.7 GHz, and timing is calculated using the Python time.clock() method, which measures processor system time in seconds. Table 4 shows the training time using scikit-learn's Gaussian process regression \texttt{fit} function and inference time using its \texttt{predict} function \citep{Pedregosa}. Inference time for obtaining the ``ground-truth'' distribution includes the inference time of the 1000 samples per test point, as well as extra computation for calculating the mean and variance of the posterior predictive distribution for each test point. Similarly, for the inverse-variance weighted approach, inference time includes the inference time of the five samples per test point as well as the extra computation. Inference times for noise-contaminated observations are averaged for white and pink noise.

\begin{table}
\centering
\caption{Fitting and Inference Times (seconds)} 
\begin{tabular}{l c c | c c c c c}
\hline
& Train/Test Ratio & No.$^a$  & Training & MC & Na\"{i}ve & IVW5$^b$ & IVW10$^c$ \\
\hline
\textbf{Case 1} & 80/20 & 62 & 1.10E+00 & 3.14E-01 & 6.56E-04 & 2.13E-03 & 7.94E-03\\
& 70/30 & 54 & 8.46E-01 & 4.14E-01 & 1.00E-03 & 4.83E-03 & 5.02E-03\\
& 60/40 & 47 & 9.09E-01 & 4.58E-01 & 1.06E-03 & 4.18E-03 & 2.12E-02\\
& 50/50 & 39 & 9.87E-01 & 4.93E-01	 & 1.32E-03 & 3.23E-03 & 1.52E-02\\
\hline
\textbf{Case 2} & 80/20 & 62 & 6.30E-01 & 2.34E-01 & 8.89E-04 & 1.72E-02 & 2.77E-03\\
& 70/30 & 54 & 3.35E-01 & 2.93E-01 & 7.49E-04 & 8.17E-03 & 1.08E-02\\
& 60/40 & 47 & 6.12E-01 & 3.59E-01	 & 8.62E-04 & 2.54E-03 & 9.14E-03\\
& 50/50 & 39 & 5.35E-01 & 3.66E-01	 & 7.12E-04 & 2.60E-03 & 9.88E-03\\
\hline
\textbf{Case 3} & 80/20 & 62 &5.79E-01 & 2.06E-01 & 7.20E-04 & 2.19E-03 & 6.20E-03 \\
& 70/30 & 54 & 3.30E-01 & 2.67E-01	 & 5.91E-04 & 1.99E-03 & 7.01E-03\\
& 60/40 & 47 & 3.37E-01 & 3.33E-01	 & 6.87E-04 & 2.55E-03 & 6.59E-03\\
& 50/50 & 39 & 3.14E-01 & 3.18E-01 & 1.34E-03 & 2.49E-03 & 1.30E-02\\
\hline
\end{tabular}
\\ $^a$No. of training points
\\ $^b$IVW5 - Inverse-variance weighting approach for 5 samples
\\ $^c$IVW10 - Inverse-variance weighting approach for 10 samples
\end{table}

Training time (offline) in GPR is relatively modest, requiring fewer than two seconds of processor system time in all cases considered. It is interesting to note that inference time depends on the number of training points, $m$, since calculating the mean of the posterior predictive distribution requires inversion of the $K(X,X) + \sigma_n^2I$ matrix, which dominates the calculation with $\mathcal{O} (m^3)$ computational complexity. Online inference takes fewer than two milliseconds for one random sample, and 18 milliseconds for five random samples and 22 milliseconds for ten random samples using the inverse-variance weighting approach. The additional time needed for the inverse-variance weighting method is modest, compared to the time required for ground truth estimation using MC sampling, which is on the order of hundreds of milliseconds. 

\section{Conclusion}
Gaussian process regression can be a powerful tool for problems limited by small datasets, such as the duct characterization problem. Unlike many supervised machine learning algorithms, including least squares and artificial neural networks, GPR is not prone to overfitting and can provide uncertainty quantification on those predictions.

We show that GPR is able to accurately predict duct height from an array of propagation factors that is sparsely sampled within a context consistent with bistatic radar systems. For noise-contaminated propagation factors, we show that the inverse-variance weighing approach generally performs better than the na\"{i}ve approach by decreasing the MSE and increasing the variance. Based on favorable model performance in the presence of severe colored noise contamination with a low train to test ratio, the first sampling case is recommended for evaporation duct height prediction. The results in this study are obtained from surrogate data with certain environmental idealizations, such as a smooth ocean surface, constant refractivity profile in range, etc. However, GPR is expected to generalize to more realistic cases under the condition that the training data reflects these less idealized conditions. From these results and recorded inference times, Gaussian process regression provides suitable, ``real-time'' approach for estimating duct height.

\section{Acknowledgments}
Datasets and code for this research are available in this in-text data citation reference: \citep{Code}. The authors gratefully acknowledge ONR Division 331 \& Dr. Steve Russell for the fianancial support of this work through grant N00014-19-1-2095.

\bibliography{paperbib}

\end{document}